\documentclass[10pt,aps,twocolumn,superscriptaddress]{revtex4-1} %% REVTeX 4.0
\pdfoutput=1
\usepackage{amsmath}
\usepackage{float}
\usepackage{graphics}
\usepackage[normalem]{ulem}
\usepackage{color}

\definecolor{Black}{RGB}{0,0,0}

\begin{document}
\title{The influence of an electric field on photodegradation and self healing in disperse orange 11 dye-doped PMMA thin films.}
\author{Benjamin Anderson, Sheng-Ting Hung, and  Mark G. Kuzyk}
\address{Department of Physics and Astronomy, Washington State University,
Pullman, WA 99164-2814}
\date{\today}

\begin{abstract}
The influence of an applied electric field on reversible photodegradation of disperse orange 11 (DO11) doped into PMMA is measured using digital imaging and conductivity measurements.  Correlations between optical imaging, which measures photodegradation and recovery, and photoconductivity enables an association to be made between the damaged fragments and their contribution to current, thus establishing that damaged fragments are charged species, or polarizable.  Hence, the decay and recovery process should be controllable with the applications of an electric field.  Indeed, we find that the dye polymer system is highly sensitive to an applied electric field, which drastically affects the decay and recovery dynamics.  We demonstrate accelerated recovery when one field polarity is applied during burning, and the opposite polarity is applied during recovery.  This work suggests that the damage threshold can be increased through electric field conditioning; and, the results are qualitatively consistent with the domain model of Ramini\cite{Ramini12.01,Ramini13.01}.  The observed behavior will provide useful input into better understanding the nature of the domains in the domain model, making it possible to design more robust materials using common polymers and molecular dopants.

\vspace{1em}
OCIS Codes:

\end{abstract}

\maketitle

\vspace{1em}

\section{Introduction}
One of the fundamental limitations of dye-doped polymeric devices is photodegradation.  Much work has been done to understand photodegradation and to design more resilient devices\cite{taylo05.01,exarh98.01,White91.01,Rabek95.01,Li84.01,Cerdan12.01,Yunus04.01,Fellows05.01,Kurian02.01,Zhang98.01,Vydra96.01,Mortazavi94.01,Sutherland96.01,Kochi91.01,Annieta,Cumpston95.01,Tanaka06.01, Albini82.01,Dubois96.01,Rahn94.01,Avnir84.01,Knobbe90.01,Kaminow72.01}.  In the late nineties a novel effect of self-healing was observed in rhodamine and pyrromethe dye-doped polymer fibers\cite{peng98.01}. Later, self-healing was reported in the anthraquinone derivative disperse orange 11 (DO11) doped into PMMA\cite{howel02.01}, as well as Air Force 455 (AF455) doped into PMMA\cite{zhu07.01,desau09.01}, \textcolor{Black}{ 8-hydroxyquinoline aluminum (Alq3)\cite{Kobrin04.01}} and other anthraquinone derivatives\cite{Anderson11.02}.   The dyes exhibiting self healing cover a wide range of structures and sizes, suggesting a general fundamental mechanism is responsible.

\textcolor{Black}{While the mechanisms of irreversible photodegradation in dye-doped polymers are not fully understood, two of the most common mechanisms considered are photo-induced reductive cleavage of bonds\cite{Zhang98.01,Vydra96.01,Mortazavi94.01} and photooxidation, in which absorbed light generates ions/free radicals which interact with the polymer chains to form singlet oxygen\cite{Rabek95.01,Sutherland96.01,Kurian02.01,Cerdan12.01} and/or electron donor-acceptor complexes\cite{Kochi91.01,Annieta} which lead to irreversible chemical reactions such as hydrogen abstraction (Norrish type II)\cite{Rabek95.01,Tanaka06.01, Albini82.01}, formation of a carbonyl species (Norrish type I)\cite{Cumpston95.01,Sutherland96.01}, formation of hydroperoxide\cite{Li84.01}, and/or other reactions\cite{Rabek95.01}.  These mechanisms are characterized by their irreversibility.}

\textcolor{Black}{Measurements of the {\em irreversible} process under the actions of an applied electric field\cite{Khan97.01,Su13.01,Kang99.01} and as a function of temperature\cite{Rabek95.01,Gonzalez99.01,Gonzalez00.02,Ramini12.01,Ramini13.01} show behavior opposite to what is observed in {\em reversible} photodegradation.  Increasing the electric field, as we show in this paper, slows the degradation process whereas irreversible photodegradation is accelerated by an electric field.  Furthermore, if the recovery process observed in DO11/PMMA is due to transients generated during photodegradation, it would be accelerated with increased temperature.  Self-healing materials exhibit a decrease in the healing rate with temperature.  This suggests that certain material systems, as reported here, behave very differently than typical dye-doped polymeric materials so that the recovery process originates from a different mechanism than is typically studied.}

\textcolor{Black}{The photodegradation time scale in self-healing polymers is similar in its dose-dependence to other dye-doped polymers\cite{Gonzalez99.01,Gonzalez00.01,Gonzalez00.02,Gonzalez01.01}, but, these other systems do not recover.   Furthermore, the degradation population dynamics and absorption spectra appear similar in liquid and polymer\cite{howel02.01,howel04.01}, but only degradation in the polymer is observed to be reversible.  These similarities in the decay process suggest that the mechanisms may be similar, but something unique in the interaction between the DO11 chromophore and the polymer fosters recovery.  Irreversibility is characterized by the requirement that the reverse process is highly improbable by virtue of the huge numbers of configurations that the system can occupy in phase space, and the fact that only an infinitesimal subset of them leads to recovery.  It thus may be that the degradation process is the same in all materials; but in self-healing materials, the polymer restricts the phase space sufficiently to make the reverse process more likely. Additionally, phase space restriction, due to the polymer host, is consistent with increased stability of organic dyes in solid matrices\cite{Dubois96.01,Rahn94.01,Avnir84.01,Knobbe90.01,Kaminow72.01}.}

Previously, molecular reorientation and diffusion \textcolor{Black}{were proposed as the decay mechanisms to explain reversible photodegradation, with degradation due to diffusion of dyes from the hot spot created by the focused laser or orientational hole burning due to the light's polarization  followed by recovery due to back diffusion or re-orientational relaxation.  However they} were eliminated as the mechanisms of self healing \textcolor{Black}{due to conclusive measurements of polarization independent recovery\cite{embaye08.01} and spatially resolved transmittance imaging of recovery that were inconsistent with diffusion\cite{ramini11.01}. Additionally, the observation of an isobestic point in the family of absorption spectra during decay\cite{embaye08.01} points to a conversion from one species to another, as one would expect in a true degradation process.}

All of these observations during recovery of anthraquinone derivatives suggest that changes in the molecular structure are associated with reversible photodegradation.  One hypothesis is that the light induces  the anthraquinone dye to convert to a tautomer state (phototautomerization), which can then bind with another tautomer to form a dimer\cite{embaye08.01}. Another proposed mechanism is that photodegradation is associated with the formation of a twisted internal charge transfer (TICT) state, and recovery occurs when the sample relaxes back to the ground state\cite{Westfall12.01}.  While phototautomerization and TICT are plausible hypotheses for anthraquinone derivatives, these mechanisms may not be able to explain reversible photodegradation observed in other dyes. \textcolor{Black}{The details of the process are immaterial for the present investigations.}

In one of the studies concerning AF455\cite{desau09.01}, the authors propose that photocharge ejection and recombination \textcolor{Black}{(differing from the photooxidation mechanism)} may be responsible for reversible photodegradation.  The basic principle is that when a dye molecule decays it ejects an ion fragment into the surrounding environment that is free to move about; it eventually returns and recombines with the damaged molecule, returning the dye to its initial state. In our current study we seek to test this hypothesis by studying the influence of an electric field during decay and recovery. If charged species are created during \textcolor{Black}{reversible} decay then an applied electric field should change the decay and recovery characteristics of the sample, leading to a measurable current due to the ejected charges.

\section{Method}
\begin{figure}
\centering
\includegraphics{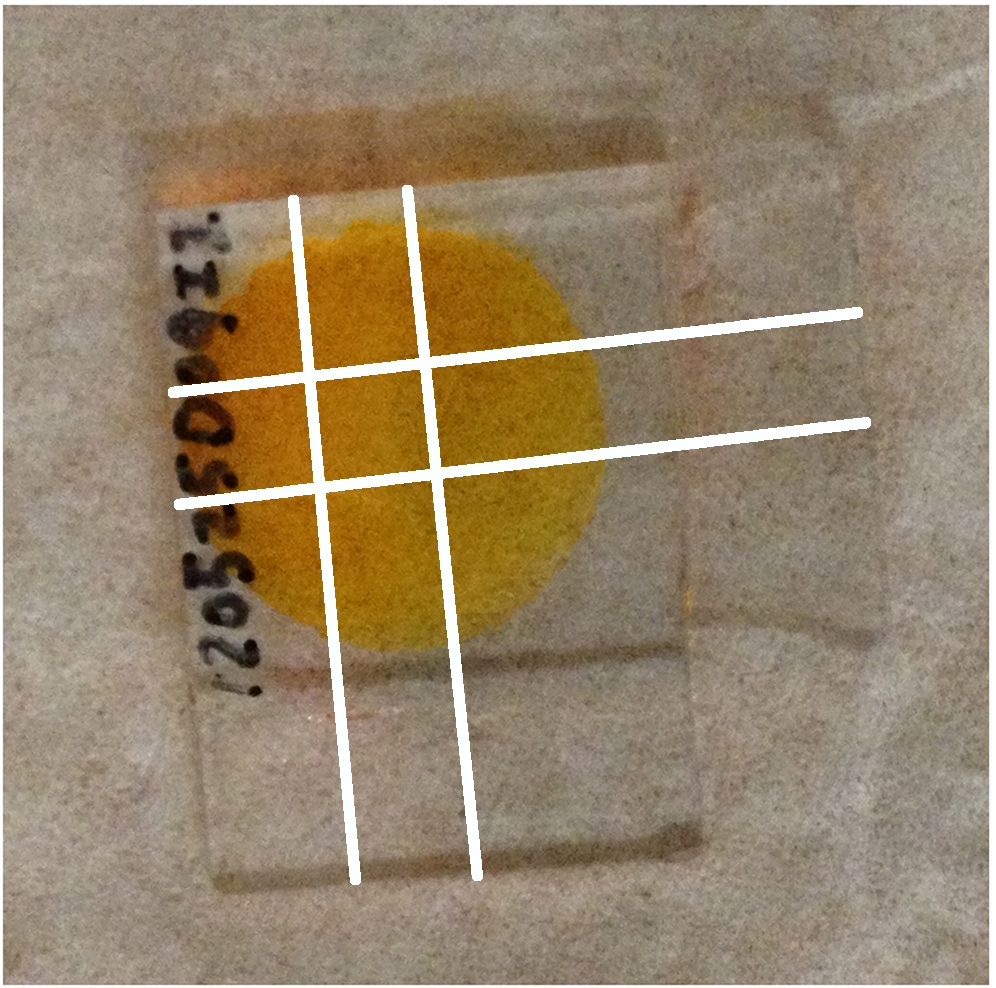}
\caption{Image of a sample made of two crossed ITO substrates with dye-doped polymer pressed in between.  Overlayed white lines mark the boundaries of the ITO strips.}
\label{Fig:Sample}
\end{figure}

The samples used in our studies are DO11 doped into PMMA sandwiched between two indium tin oxide (ITO) coated glass slides.  The ITO glass substrates are etched using an aqueous solution of 20\% HCl heated to 50$^\circ$C for 20 min, then cut into 25 $\times$ 19 mm rectangles.  To prepare the dye-doped polymer; PMMA and DO11 from Sigma-Aldrich are dissolved in $\gamma$-butyrolactone and propylene glycol methyl ether acetate (PGMEA).  The solutions are stirred with magnetic stirrers and filtered through a 0.2$\mu$m syringe filter to remove particulates.  Once the solution is prepared, the cut and etched ITO glass substrates are placed on a hot plate at 50$^\circ$C, and several drops of the solution are carefully placed on the ITO side of the substrates.  The hot plate's temperature is then raised to 95$^\circ$C to induce solvent evaporation.  The hot glass substrate and dye-doped polymer are then transferred to a vacuum oven overnight to ensure the samples are dry.  The glass-polymer sample is then combined with a second ITO glass substrate to form a sandwich structure, as shown in Figure \ref{Fig:Sample}.  The sandwich structure is pressed for 135 min at a pressure of $72$psi and a temperature of 130$^\circ$C.

\begin{figure}
\centering
\includegraphics{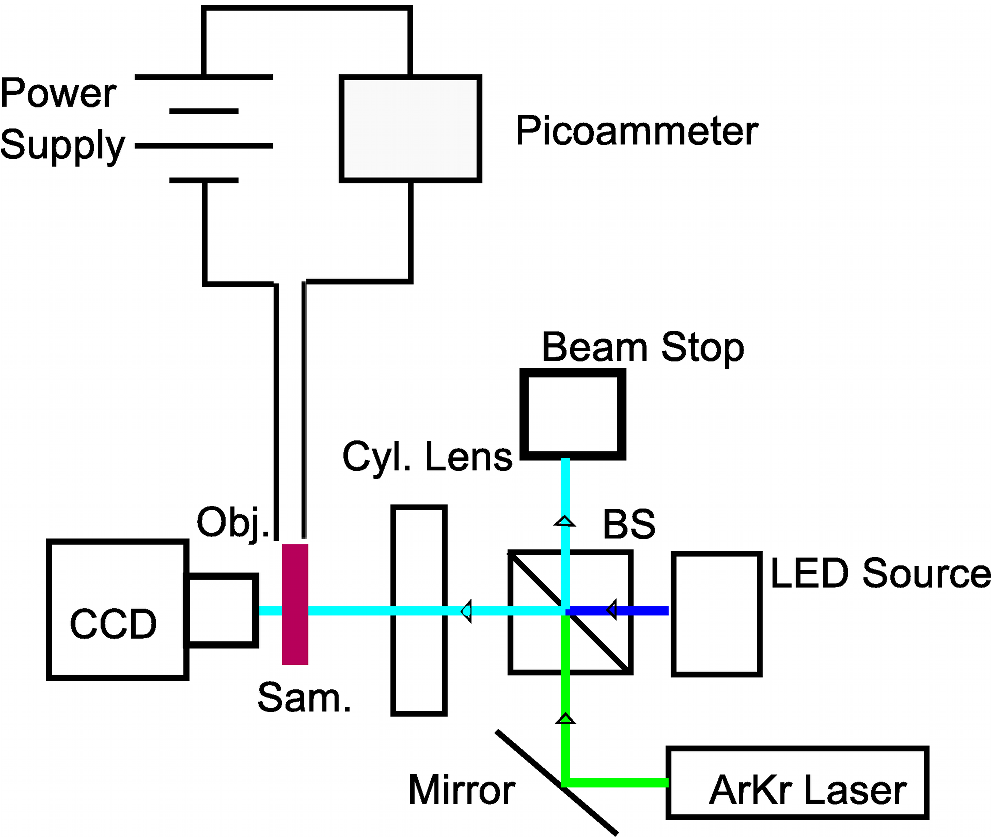}
\caption{The experiment. An ArKr laser burns the sample, a LED illuminates it and the CCD detector images the burn.  The sample is connected in series with a power supply and pico ammeter that measures the current through the sample.}
\label{Fig:Apparatus}
\end{figure}

During measurements, the samples are held in place using an acrylic sample holder with plastic screws to avoid electrical contact with the ITO conductor.  An electric field is applied to the polymer via a potential difference between the two ITO strips using an SRS PS 350 DC high voltage power supply in series with an RBD picoammeter to measure current through the sample.  We measure optical decay and recovery using a digital imaging microscope\cite{Anderson11.01,Anderson11.02,Anderson12.01} while measuring the current flowing through the sample.   A schematic of our apparatus is shown in Figure \ref{Fig:Apparatus}.

We find a dark current response to a step function voltage to decay as a multiple exponential.  After a long period of time, typically hours, the dark current approaches a ``steady state''. In the steady state, all measured changes of current depend only on the influence of the pump beam.

The experimental protocol is as follows. First, the voltage source is turned on and the current reaches the steady state (usually 2 hrs).  An ArKr pump laser operating at 488nm is focused to a line on the sample  for a peak intensity of 175 W/cm$^2$.  The transmittance of the sample under illumination of a custom LED source is measured with a micro-imaging camera to probe burning.  Shutters are used to block the pump beam for several seconds to protect the camera during imaging.  During recovery the pump is blocked and images are taken at semi-log time intervals (1min, 10min, 30min).

We define the $y$-axis to be along the burn line and the $x$-axis to be perpendicular to it, with $x=y=0$ at the center of the burn.  Image line profiles are measured parallel to the $x$-axis at each $y$ coordinate.  To reduce the error due to  statistical fluctuations in the camera, ten adjacent scans are averaged to obtain a mean line profile.  The line profiles are fit to a Gaussian function to determine the amplitude, $A$, and width, $\sigma$, of each profile.  Using the fit parameters $A$ and $\sigma$ we then calculate the change in transmission ($\Delta T(x,y,t)$) at 21 $x$ positions along the profile.

Using $\Delta T(x,y,t)$ we can compute the damaged population as a function of position and time
\begin{equation}
n(x,y,t)=\frac{1}{\Delta\sigma L}\ln[\Delta T(x,y,t)],
\end{equation}
where $L$ is the sample thickness, and $\Delta\sigma$ is the difference in absorption cross section between the damaged and undamaged molecules.  In practice $\Delta\sigma L$ varies across samples as they are not perfectly uniform or homogenous.  Therefore we define a new quantity which we call the scaled damage population
\begin{eqnarray}
n'(x,y,t)&=&\Delta\sigma L n(x,y,t)
\\ &=&\ln[\Delta T(x,y,t)].
\end{eqnarray}

Using our previous two-level model for decay and recovery\cite{embaye08.01} with the addition of an irreversibly damaged component\cite{Anderson11.02}, we can write the temporal dependence of the scaled damaged population for decay and recovery as
\begin{eqnarray}
n'(x,y,t)=n_0'\left(1-e^{-\gamma t} \right), \label{Eqn:nd}
\\ n'(x,y,t)=n_{IR}+n_Re^{-\beta t}, \label{Eqn:nr}
\end{eqnarray}
where $\gamma=\beta+\alpha I$ according to our model of self healing\cite{embaye08.01,Anderson11.02}.  $n_{IR}$ and $n_R$ are constants that depend on the scaled damaged population when the laser is turned off to start the healing process.  \textcolor{Black}{$n_{R}$ is the portion of the scaled damaged population which recovers} and $n_{IR}$ is the scaled population in the limit of infinite healing time.

\section{Results and discussion}
\subsection{Dark conductivity}
\begin{figure}
\centering
\includegraphics{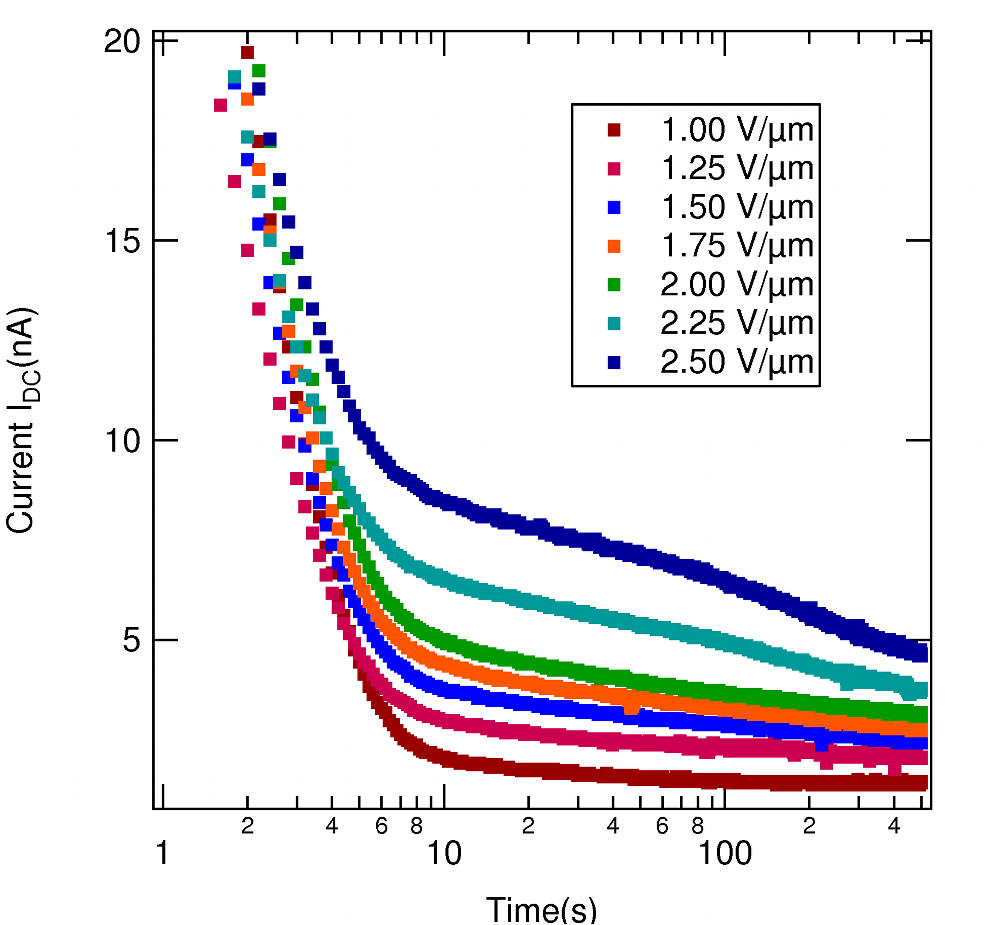}
\caption{Dark current as a function of time after the field is applied for various field strengths for an 11.5g/l sample of DO11 in PMMA polymer.}
\label{Fig:DC}
\end{figure}

 In general the initial application of an electric field results in a large sharp peak in the current, which is followed by multiple-exponential decay, the longest of which is longer than experimental time scales as shown in Figure \ref{Fig:DC}.  Typically the sample's current settles to a semi-constant level after several hours of having a field applied, after which the current takes several days to change noticeably.  This result is consistent with previous measurements of \textcolor{Black}{dark conductivity in} dye-doped polymers\cite{zimme94.01,Chilton95.01,Vijayashree92.01,Yadav93.01,Shepherd10.01,Son10.01,Zan08.01,Garcia09.01,Fukushima98.01,Ieda84,Sisk95.01,Yakuphanoglu08.01}.

 The accepted explanation for the transient dark current is \textcolor{Black}{as follows.} Initially, when the field is applied, free charges in the polymer quickly diffuse away, leading to a large current spike, which settles after the free charge is depleted.  Subsequently, several slower processes of lower magnitude -- such as alignment of polymer chains and dopants, trapped charge stripping, thickness change due to the electromechanical effect, dielectric constant changes due to the applied field, and temperature change due to ohmic heating -- are observed.

Optical absorption directly probes the population of chromophores and differentiates between species by characteristics of the spectrum.  The aim of the present work is to correlate the population of damaged and undamaged molecules with current to help gain a better understanding of the properties of the charged species that are responsible.

\subsection{Photoconductivity}
After dark conductivity measurements the sample's photoconductivity is measured during decay and recovery.  The results from the photoconductivity measurements are difficult to relate to the damaged population as many different processes are involved,  \textcolor{Black}{some} of which have similar time scales to population damage, making a distinction between them difficult.

\begin{figure}[h!]
\centering
\includegraphics{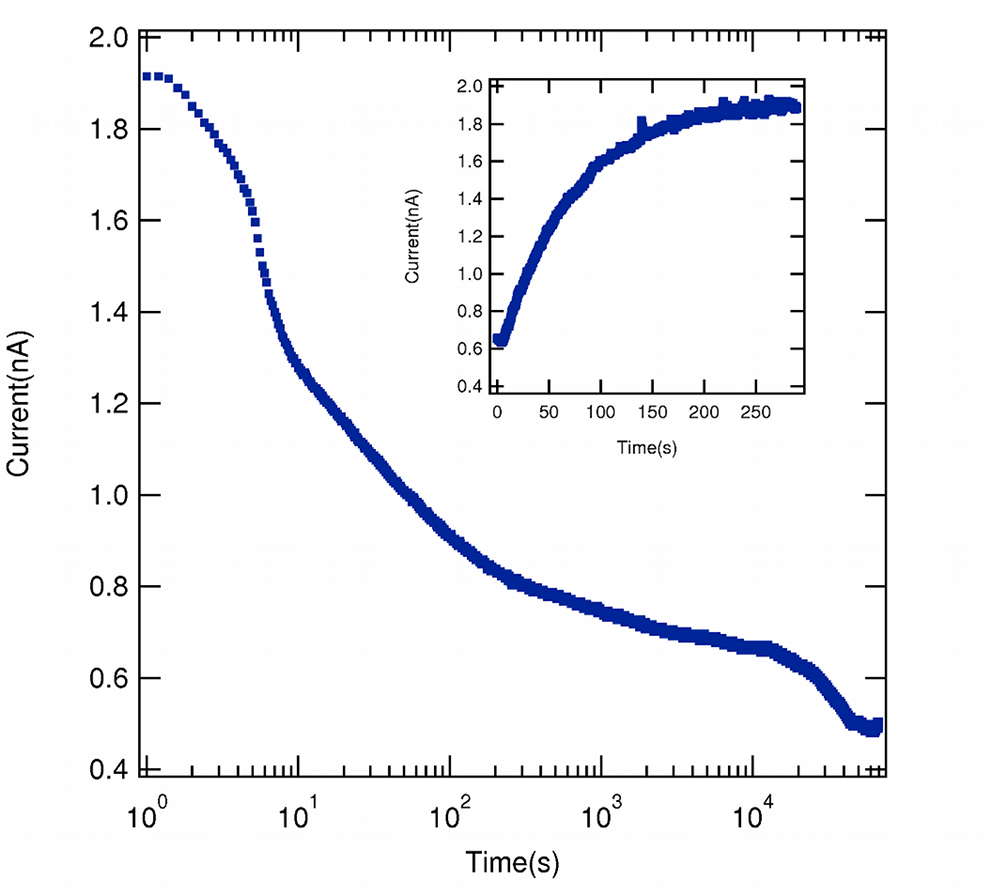}
\caption{Example of current relaxation after light exposure.  The relaxation process is characterized by multiple time scales.  Inset:  Current during light exposure follows a double exponential.}
\label{Fig:PCrisefall}
\end{figure}

Figure \ref{Fig:PCrisefall} shows both the double exponential response of the current to the pump light being turned on (inset) as well as relaxation when the light is turned off.  The ratio of fast and slow exponential rate components is typically 10, with the fast component having the larger amplitude. Optical absorbance measurements during pumping match the rate of the slow component of the current response.  At lower intensities where optical damage is negligible, the fast component of the photocurrent is observed, but the slow component is negligible.

 We hypothesize that the fast component corresponds to a variety of processes such as molecular reorientation, non-damaging photocharge ejection, and excitation of trapped charges; while the slow component is primarily due to damaging photocharge ejection and processes involving the reorientation and diffusion of polymer chains.

The relaxation of the current is a complex process involving multiple characteristic time-scales, with the current relaxing back to almost the initial current level in several minutes.  After tens of minutes the current returns to the pre-illumination level, and after several hours the current decreases below the initial value.  Given that for both decay and recovery the fast response is also the largest response, we hypothesize it is due to recombination of ejected charges and charges becoming trapped.  The long time-scale relaxation rate of optically probed recovery and transient dark current matches; therefore, we postulate that the process involved is recombination of charged fragments made in the damage process and the slow movement of polymer chains.

\begin{figure}[h!]
\centering
\includegraphics{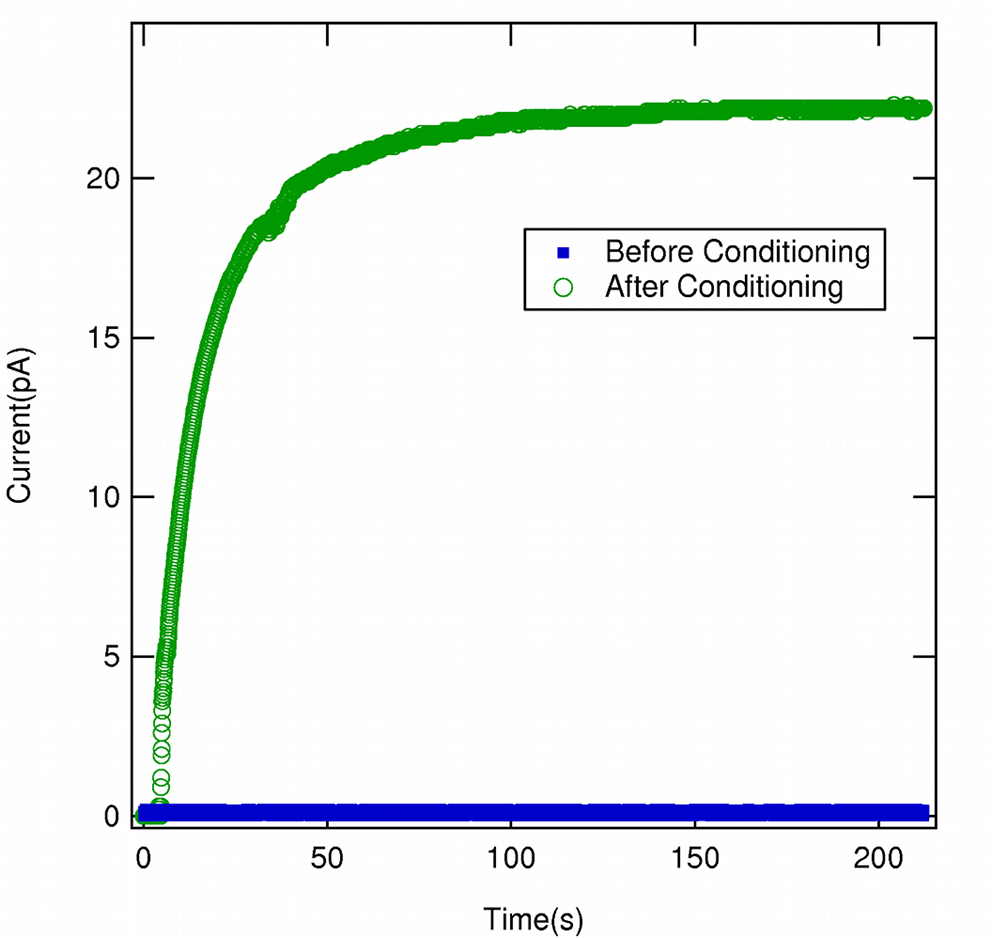}
\caption{Measured zero-field photocurrent before and after electric field conditioning for one week.}
\label{Fig:0VPC}
\end{figure}

Initial studies characterized the photocurrent with no electric field applied to the sample. All fresh samples display no zero-field photocurrent.  After placing the samples in electric fields for several weeks, a zero-field photocurrent is observed as shown in Figure \ref{Fig:0VPC}.  This photocurrent is almost 100 times smaller than with an applied field (20 pA versus 2 nA) and persists for over one week.  After a period of relaxation the zero-field photocurrent returns to a negligible level.

The observation of a zero-field photocurrent, after the \textcolor{Black}{conditioning} field had been applied, suggests the presence of a quasi-stable polarization field which persists even after the applied field is turned off. Counterintuitively though, we find the zero-field photocurrent to be in the same direction as the field applied during conditioning.  Typically, induced electric fields in dielectrics are antiparallel to the applied field, either due to polarization effects or moving charges.  Yet our observations are contrary to our expectations and currently we have not found a suitable explanation.  Note that this result is not used in the analysis of the data that follows.

\subsection{Electric field dependent decay and recovery}
Digital imaging microscopy is used to measure decay and recovery simultaneously with photoconductivity to test the effects of an applied electric field. Samples of 7g/l, 9g/l, and 11.5g/l were studied.  Our results for all samples are qualitatively consistent, but with large variability.  While the temperature and pressure are kept constant during sample fabrication, we believe variations in humidity and solvent evaporation rate may have played a large role in sample-to-sample variability.  Thus, we only use qualitative results in the analysis of our observations.

\textcolor{Black}{We consider} three characteristic time periods, \textcolor{Black}{which} are: (1) before electric field conditioning, (2) during conditioning where the field is on for one to two weeks, and (3) after conditioning.  Our decay and recovery results are noticeably different during each period.

\begin{figure}
\centering
\includegraphics{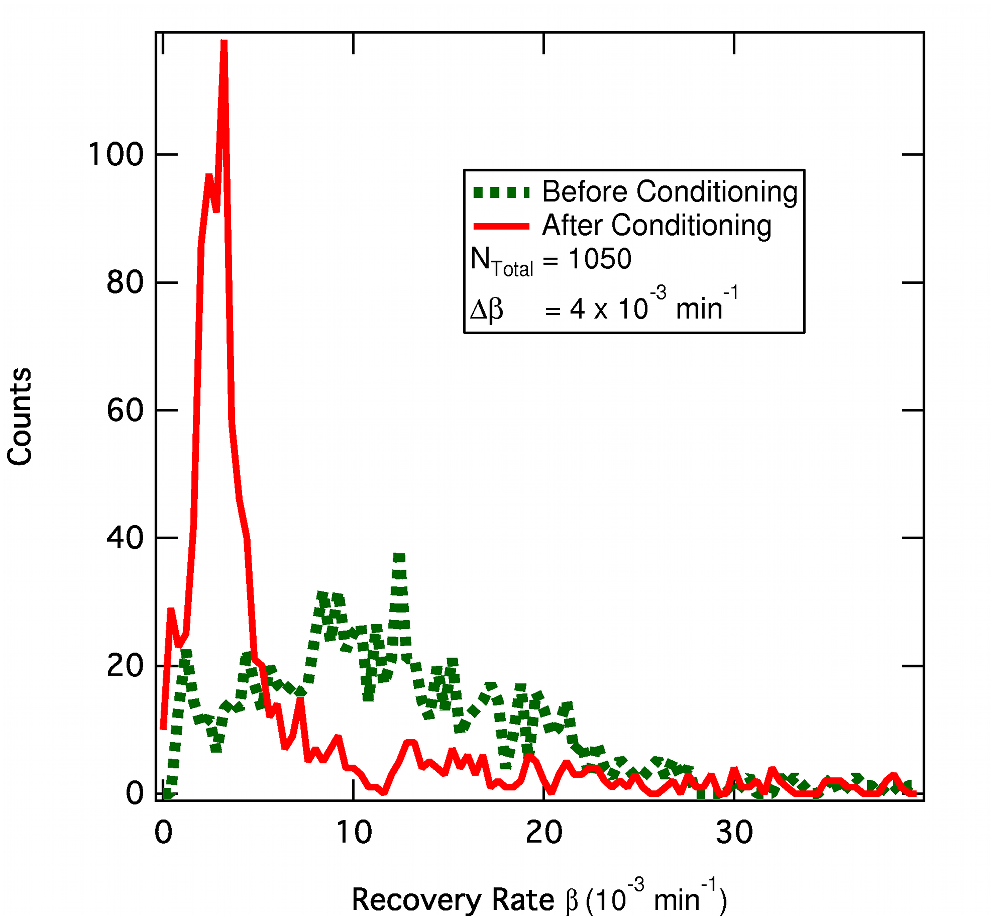}
\caption{Zero-field recovery rate distribution for a 7g/l sample before and after  conditioning with an electric field. Conditioning leads to a narrower distribution}
\label{Fig:0Vhist}
\end{figure}

The most pronounced effect of electric field conditioning is found from the distribution of zero-field recovery rates, which we determine using over 1000 points on a sample with optical transmittance imaging as a probe.  Figure \ref{Fig:0Vhist} shows the histogram of zero field recovery rates for a 7g/l sample before and after electric field conditioning.  Note that after conditioning the histogram is much narrower and well defined.  The post-conditioned peak corresponds to a lower recovery rate than the pre-conditioned case.  However, we \sout{have} also observe the opposite behavior, where the post-conditioned peak corresponds to a higher recovery rate.  There appears to be no correlation between the post-conditioned peak and the pre-conditioned peak, but in all cases the post-conditioned peak is drastically narrower.

We also study decay and recovery of the samples during conditioning with a voltage applied and found that the electric field has a distinct effect on the photodamage and healing process.  During treatment, two identical measurements on the same sample separated by several days may yield drastically different results.  The fact that our results during treatment were dependent on how long the field had been applied, along with the large variability of the pre-conditioned sample, and the reduction in variability after conditioning leads us to hypothesize that trapped and free charges, along with the alignment of the polymer chains play a crucial role in the mechanism behind recovery.  Electric field conditioning essentially smooths out the sample's trapped/free charge density and aligns the polymer chains such that the \textcolor{Black}{charge densities at recovery sites are made more uniform.} 

\begin{figure}[h!]
\centering
\includegraphics{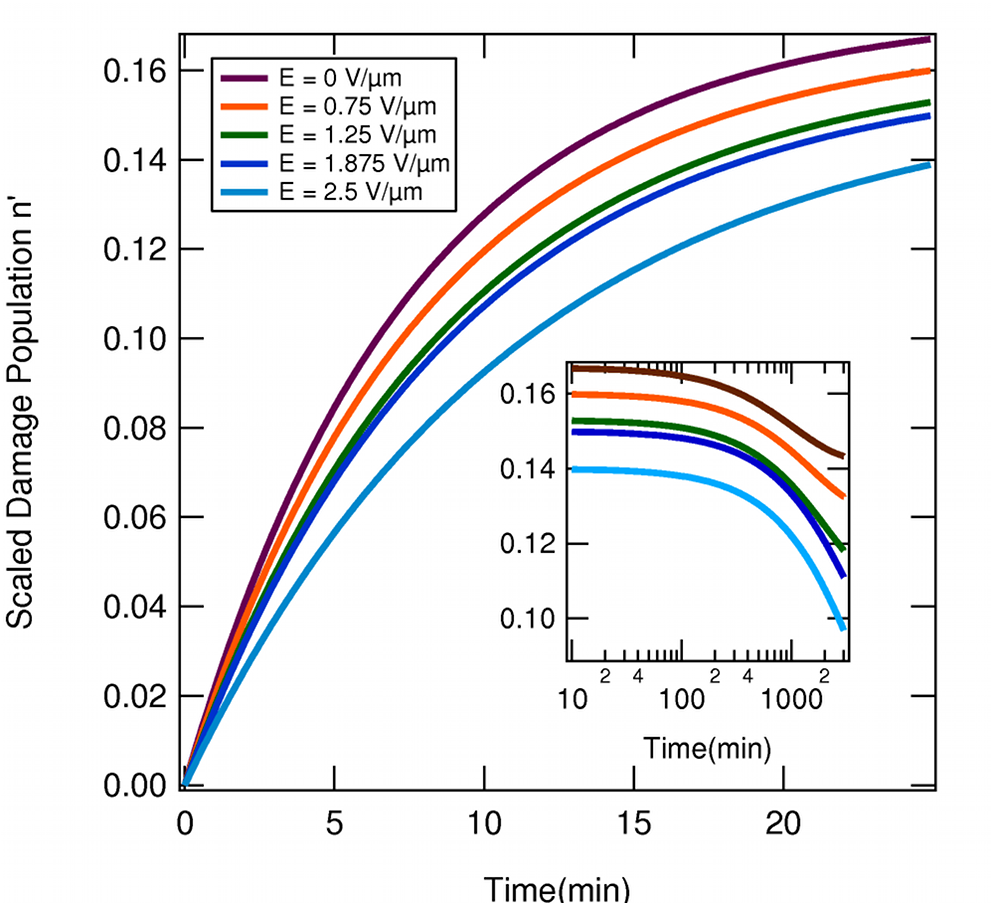}
\caption{Scaled damaged population as a function of time as measured with optical absorption imaging at the center of the burn ($x=y=0$) for various applied fields during photodegradation.  Inset:  Self healing at the same spot and applied field.  The curves shown are fits to the data which are not shown for clarity.}
\label{Fig:FieldDIM}
\end{figure}

After conditioning we find an applied electric field has a repeatable effect on a sample's decay and recovery rate.  Figure \ref{Fig:FieldDIM} shows the scaled damage population curve obtained by a fit to the optical imaging data for various applied fields for decay and recovery (inset) for a 9g/l sample.  The data in Figure \ref{Fig:FieldDIM} is for a single polarity of the applied field for both decay and recovery.  There is a noticeable change in the degree of decay and recovery as the applied field is increased.  This effect is found to be universal across all conditioned samples, albeit differing in scale.

\begin{figure}
\centering
\includegraphics{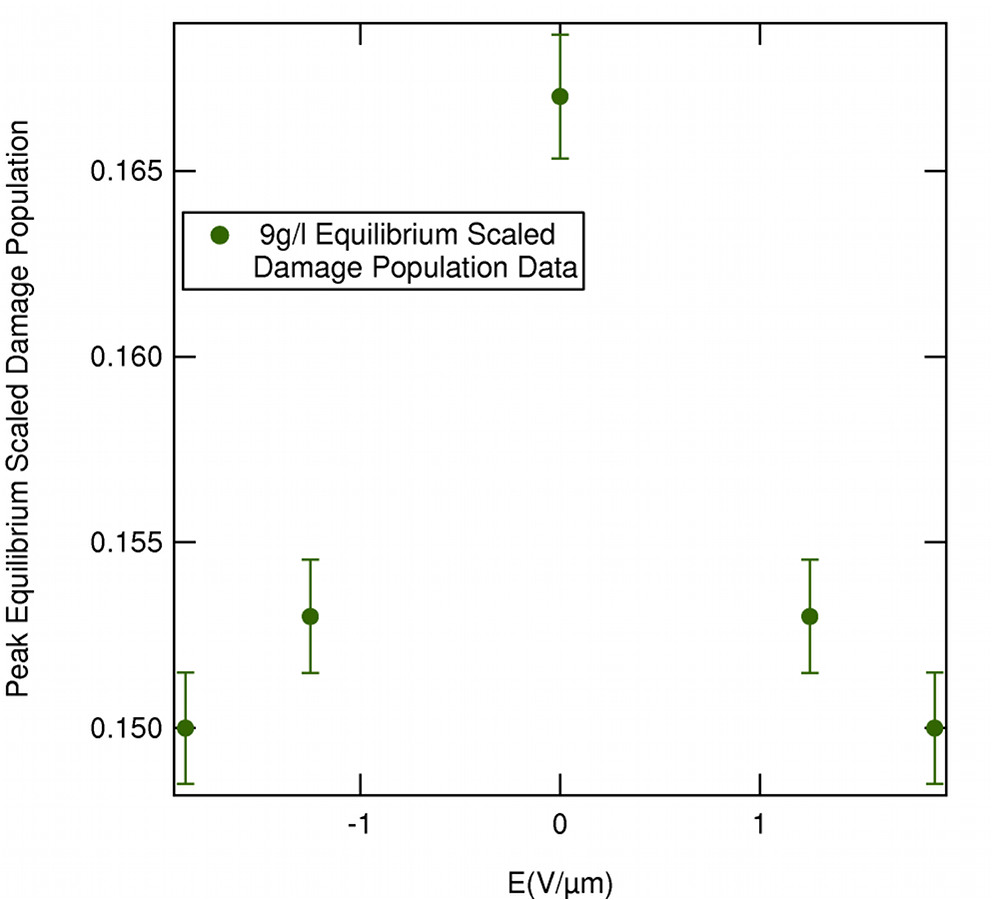}
\caption{Equilibrium scaled damaged population for the center of the beam ($x=y=0$).  As the field strength is increased the degree of damage decreases regardless of polarity.}
\label{Fig:dp}
\end{figure}

\begin{figure}
\centering
\includegraphics{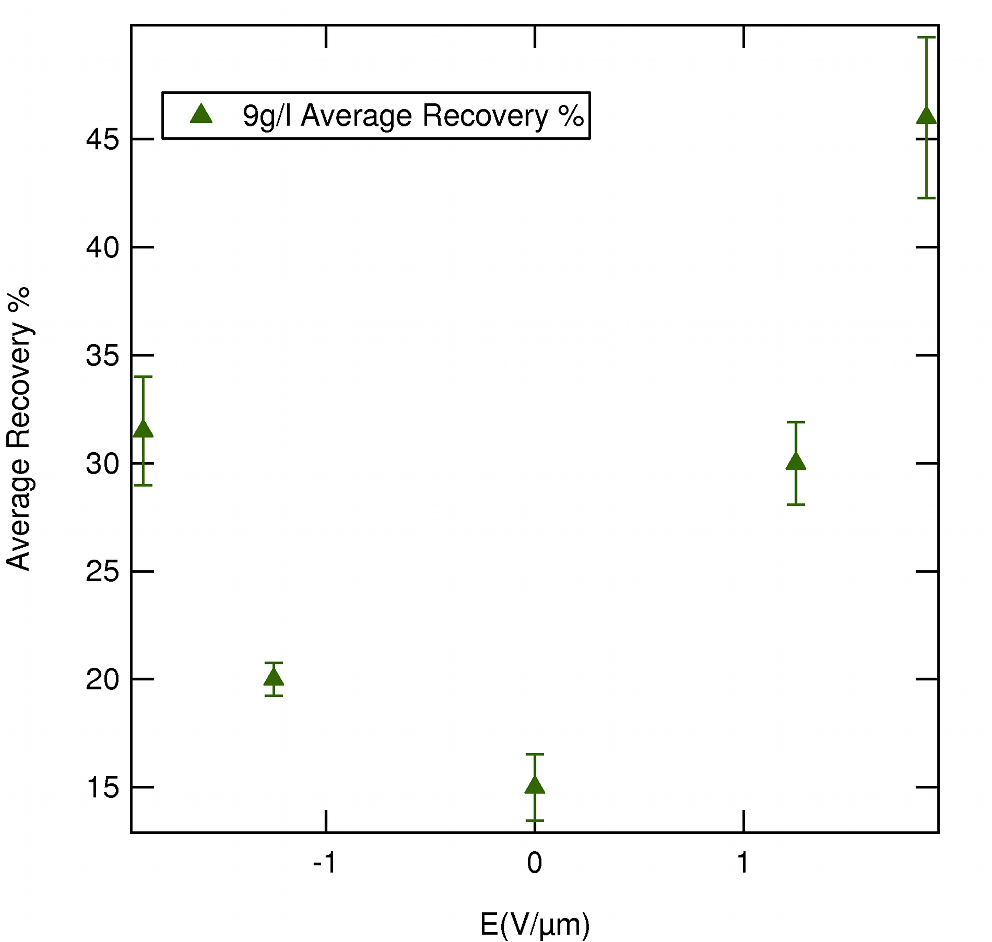}
\caption{Average recovery fraction as a function of applied electric field.  A negative electric field points in the opposite direction to the field applied during photodegradation.}
\label{Fig:rp}
\end{figure}

\textcolor{Black}{To quantify the effect of an applied electric field} the intensity independent decay rate, $\alpha$, the recovery rate, $\beta$, the peak equilibrium scaled damaged population while burning, $n_0'$, and the recovery fraction, $n_f$, which is derived from Equation \ref{Eqn:nr},
\begin{equation}
n_f=\frac{n_R}{n_{IR}+n_R},
\end{equation}
are determined.  We find the applied field, independent of direction, mitigates damage and increases the degree of recovery.  Figure \ref{Fig:dp} shows the peak equilibrium scaled damage population as a function of applied field, and Figure \ref{Fig:rp} shows the average recovery fraction.

\begin{figure}
\centering
\includegraphics{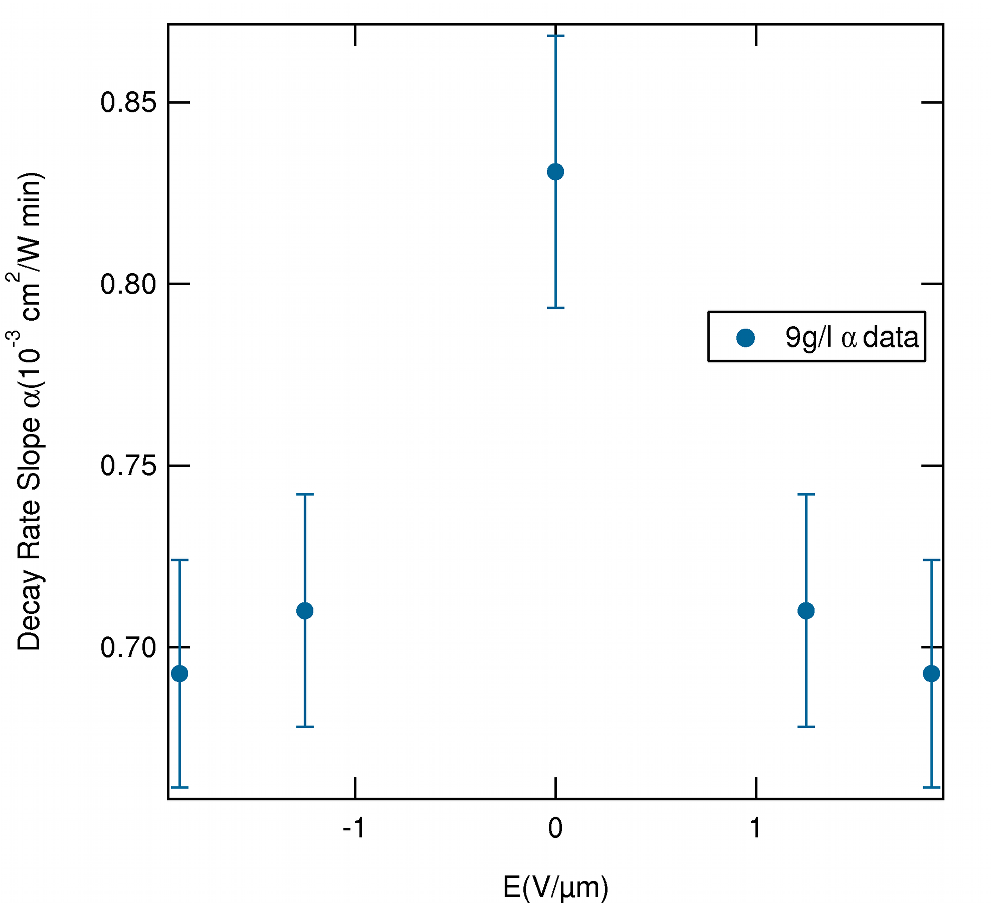}
\caption{The intensity independent decay rate, $\alpha$, as a function of applied electric field.  A negative electric field points anti-parallel to the \textcolor{Black}{pump Poynting} vector.}
\label{Fig:alphaE}
\end{figure}

The rate at which the sample decays and the rate at which it recovers is given by $\gamma$ and $\beta$\textcolor{Black}{, respectively}.  $\gamma=\alpha I + \beta$ depends on the pump intensity, $I$, so its slope gives the intensity independent decay rate, $\alpha$. Varying the applied electric field's magnitude yields $\alpha$ as a function of field strength.  Figure \ref{Fig:alphaE} shows $\alpha$ as a function of the field where positive values correspond to \textcolor{black}{the applied electric field being parallel to the the} pump Poynting vector.  The change in $\alpha$ with applied field is symmetric in both directions within experimental uncertainty.

\begin{figure}
\centering
\includegraphics{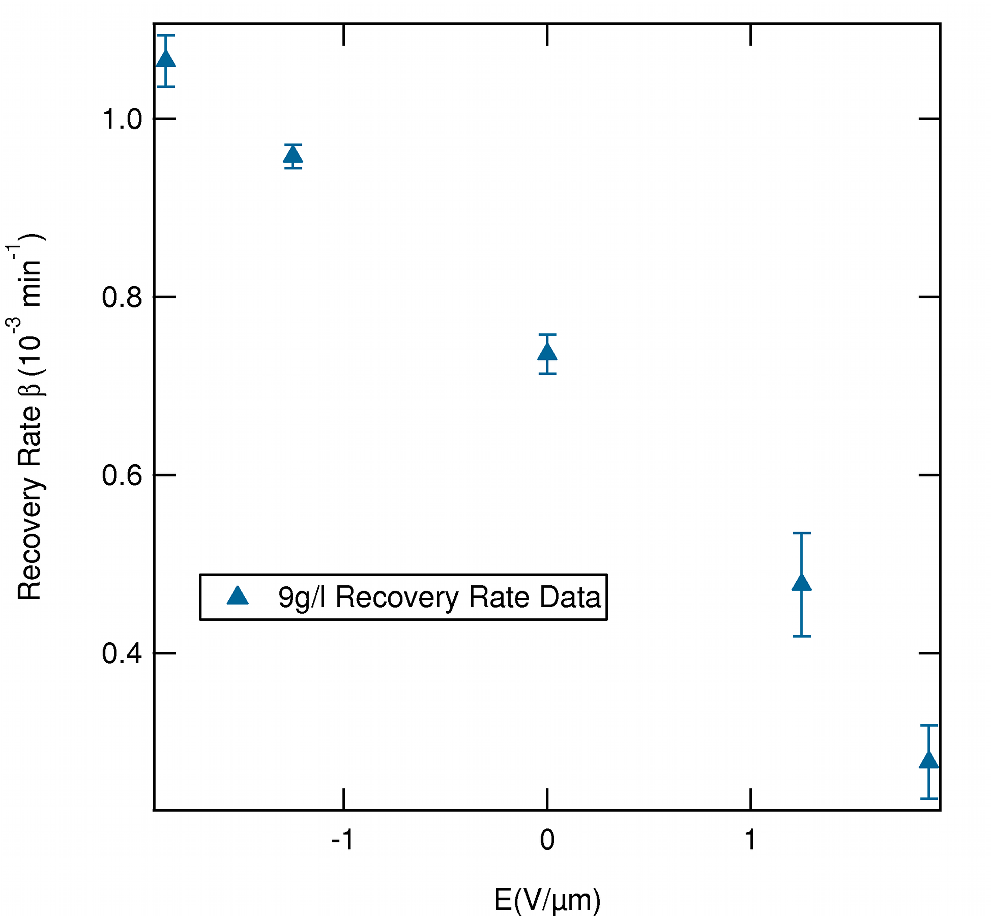}
\caption{The recovery rate, $\beta$, as a function of applied electric field. \textcolor{black}{A negative electric field points in the opposite direction to the field applied during photodegradation.}}
\label{Fig:betaE}
\end{figure}

Figure \ref{Fig:betaE}, shows that the recovery rate depends not only on the magnitude of the applied field but also its direction. If the field applied \textcolor{Black}{during recovery is} parallel to the field applied during decay, the recovery rate decreases, whereas when \textcolor{Black}{the field during recovery is} applied antiparallel, the recovery rate increases.

Increasing the field strength can \textcolor{Black}{enhance} healing. While this phenomena has been observed in several different samples we will focus on a 9g/l sample.  When burned in the presence of an applied voltage of 30V, the sample recovers in the same field by 20\%.  When \textcolor{Black}{the voltage is} increased to 50V, then 75V, and finally 100V, the degree of recovery measured using optical imaging is observed to increase.

\begin{figure}
\centering
\includegraphics{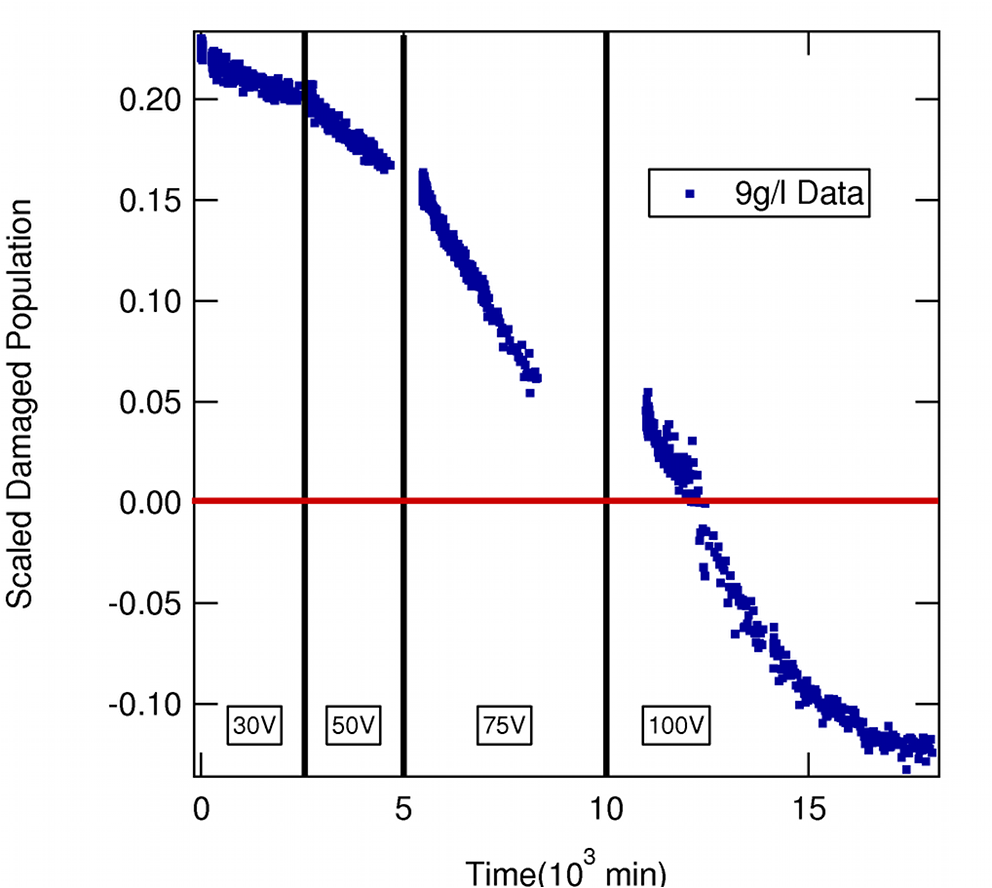}
\caption{After initially burning a sample in the presence of an applied voltage of $\Delta V$ = 30V, the recovery rate and degree of recovery increases as the voltage is increased.}
\label{Fig:Vrec}
\end{figure}

 \textcolor{Black}{Surprisingly, when the applied voltage is increased to} 100 volts \sout{applied is} \textcolor{Black}{the degree of recovery is} greater than 100\%.  Figure \ref{Fig:Vrec} shows the scaled damaged population as a function of time during which the applied field is increased as shown.

\begin{figure}
\centering
\includegraphics{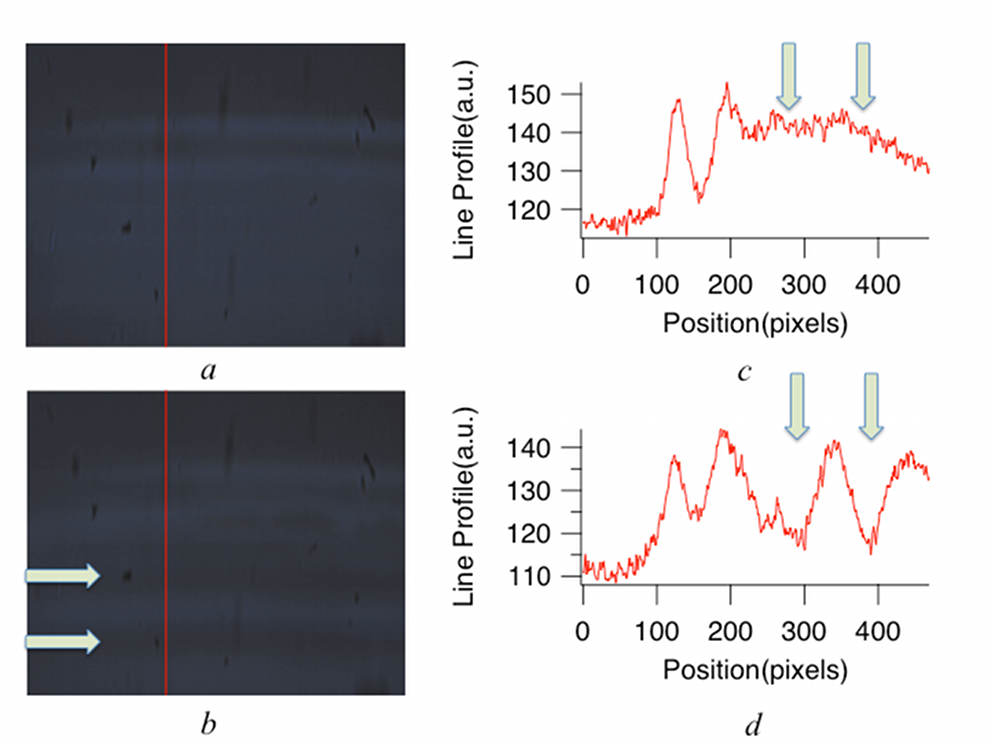}
\caption{\textit{a}: Image of horizontal burn lines when 100V is first applied (red line shows the location where the burn profile is measured).  Two of the burn lines had recovered nearly 100\%. \textit{b}: Image of burn lines after several days of 100V conditioning.  The two burn lines (marked by arrows), which had recovered to the background level, continued to recover leading to two dark lines. \textit{c}: The image line profile corresponding to the red line in \textit{a}. \textit{d}: The image profile corresponding to the red line in \textit{b}. }
\label{Fig:NegLines}
\end{figure}

Greater then 100\% recovery can be understood as follows. Figure \ref{Fig:NegLines}, shows a typical burn line that is characterized by an increase in transmittance (bright line), while the greater than 100\% recovery is characterized by a decrease in transmittance (dark line). Initially  the burn line appears bright, and as time progresses and the field is increased the line dims, eventually blending with the background, and then becoming darker than its surroundings.

\begin{figure}
\centering
\includegraphics{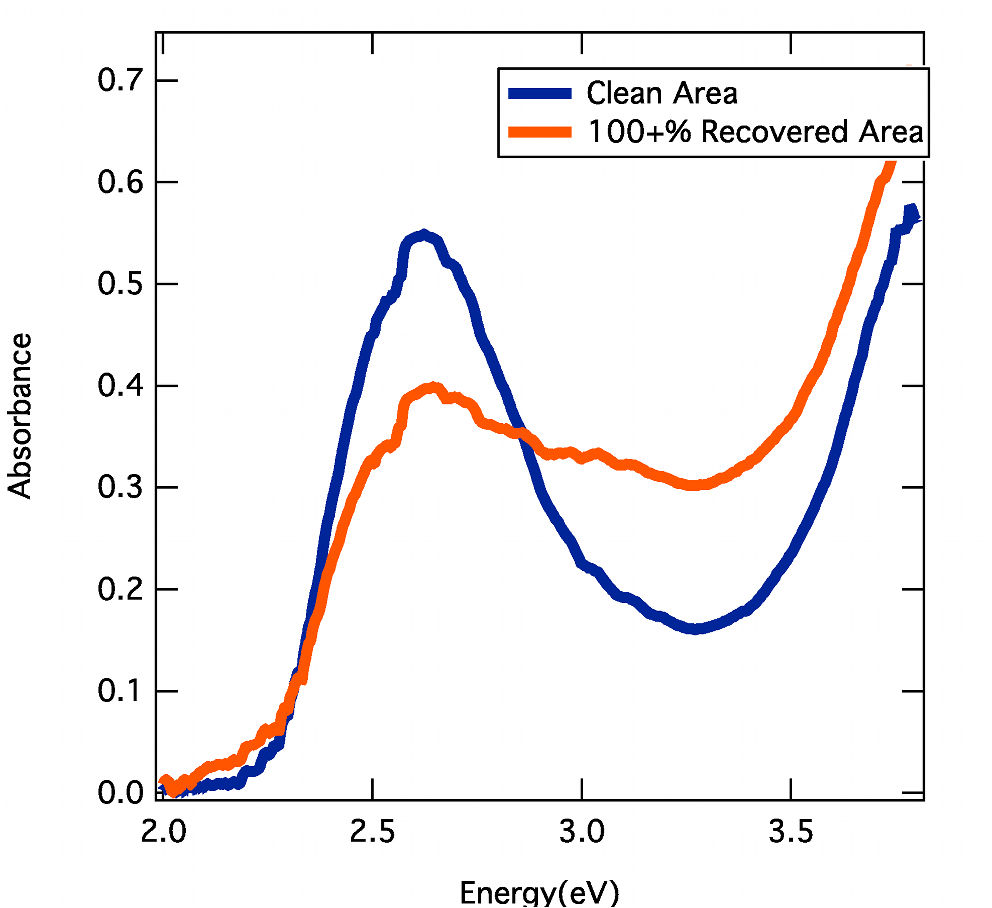}
\caption{Absorption spectrum for an undamaged sample and for area where the DIM gives 100+\% recovery.  Note that the absorbance in the deep blue region is greater than in the clean sample.}
\label{Fig:absspec}
\end{figure}

This observation suggests that the two population model is incomplete because another process appears to contribute at the time of full recovery that makes the burn line image darker than it was before burning.  Since the camera measures a mixtures of colors, it is difficult to ascertain the precise spectral change at 100+\% recovery.  In order to gain a better understanding, we use an Ocean Optics spectrometer to measure the absorbance spectrum of the burn lines that apparently recovered beyond 100\% as shown in Figure \ref{Fig:absspec}.

The absorption spectrum shows increased absorbance in the blue region of the spectrum in the 100+\% recovered area, while the peak near 2.6eV, which characterizes the non-decayed molecule has not fully recovered.  Note that the peak at 2.6eV decreases during photodamage while a new peak at 3.25eV grows and is characteristic of the reversibly damaged species.  In an ideal \textcolor{Black}{two population} recovery process, when the peak near 2.6eV recovers and becomes larger, the peak near 3.25eV should decrease in proportion to the change in the other peak's height.  However, in these experiments, we observe that the 2.6eV peak recovers and the 3.25eV peak remains constant.  The increase of the peak height in the far blue part of the spectrum suggests that a damaged species of smaller size is formed that may not be reversible.  The growth in the peak adds to the absorbance measured by the imaging system, which does not differentiate between spectral shape, thus overestimating the degree of recovery.

The nature of the irreversibly-damaged population is currently unknown, though it appears to be a distinct damaged species formed in the breakup of DO11.  Alternatively, the new peak may originate in damage to the polymer due to thermal effects induced by energy deposition into the polymer that is mediated by light absorption by the dye.  Alternatively, photodamage during the application of an electric field may lead to charged species that cause irreversible chemical reactions with the polymer or between fragments.

A domain model of self healing proposed by Ramini and coworkers\cite{Ramini12.01} successfully describes the population dynamics as measured with amplified spontaneous emission; and, a recent generalization of this theory is shown to describe a broader range of experimental conditions\cite{Ramini13.01}. While the nature of the domains of DO11 molecules is not well understood, it appears that interactions between groups of molecules, that are associated with each other, protects each group member from photodamage in proportion to the size of the domain, and accelerates molecular healing.

If the applied electric field induces larger domains, this would explain the decrease in the photodegradation rate.  However, the electric field also has the affect of sweeping the charged fragments away from a domain when a molecule is damaged.  Thus, the recovery rate should decrease if the field is applied in the same direction as during damage, and increase if the field is applied in the opposite direction, in effect forcing the charged fragments back to a domain to induce healing.  These observations show that charged species are involved in the decay and recovery process, and provide information that will be essential in building a model of the mechanisms involved.

\section{Conclusions}

Our measurements show that the decay and recovery dynamics are strongly influenced by an applied electric field, most likely due to free charge, trapped charge and polarization effects that influence the diffusion of the decay products.  The electric field appears to condition the system into a more ordered state, as evidenced by recovery measurements and the observation of an apparent quasi-stable internal electric field. Thus, applying an electric field during decay and recovery leads to a mitigation of decay and increased recovery, suggesting that with greater understanding it may be possible to design dye-doped polymeric devices which utilize electric fields to  \textcolor{Black}{increase the material's laser} damage threshold, \textcolor{Black}{as has been reported using laser cycling.\cite{kuzyk07.02}}

We propose two possible mechanisms for electric field dependent reversible photodegradation: photocharge ejection and recombination; and, reversible photodegradation linked to charged/polarizable domains.  The domain picture was recently posited as an explanation of self-healing mediated by cooperative effects between chromophores as measured with amplified spontaneous emission(ASE)\cite{Ramini12.01,Ramini13.01}.  If the domains are charged and/or polarizable, the application of an electric field would influence the decay and recovery process.  Currently, though, simple extensions to Ramini's model have failed to accurately predict the effects we observe when applying an electric field.  On the other hand, simple calculations concerning the effect of an applied field on photo ejected charges roughly match the magnitude of the effects we have observed.

We are planning on new electric field experiments varying temperature, humidity, polymer, and dye type in order to better characterize the process, which is being modified by the external field.  Generalized models that include the effect of charge generation and recombination are also being developed, where preliminary results suggest that domains of polarizable aggregates are indeed responsible.

\section{Acknowledgments}
We would like to thank Elizabeth Bernhardt for helpful input.  We also thank Wright Patterson Air Force Base and Air Force Office of Scientific Research (FA9550- 10-1-0286) for their continued support of this research.

\bibliographystyle{natbib}
\bibliography{Manuscript}

\end{document}